\documentclass[reprint, amsmath,amssymb,aps,longbibliography]{revtex4-1}
\usepackage{graphicx}
\usepackage{amsthm}   
\usepackage{kpfonts}
\usepackage[T1]{fontenc}
\usepackage[latin9]{inputenc}
\usepackage{mathrsfs}
\usepackage{bm}
 \usepackage[bottom]{footmisc}
\usepackage{amsmath} 

\usepackage{amssymb}  
\usepackage{mathrsfs} 
\usepackage{stmaryrd} 
\usepackage{mathtools}
\usepackage{color}  
\RequirePackage[dvipsnames,usenames]{xcolor}

\edef\ordinarycolon{\mathchar\the\mathcode`: }
\edef\ordinaryequals{\mathchar\the\mathcode`= }

\usepackage{breqn}
\catcode`^=7

\AtBeginDocument{%
  \catcode`^=12
}

\makeatletter

\allowdisplaybreaks

\let\cat@comma@active\@empty

\usepackage[capitalize]{cleveref}

\newif\ifnotes
\notestrue

\ifnotes
 \newcommand{\dhwc}[1]{{\color{Red}{\bf{DHW COMMENT: #1}}}}   

 \newcommand{\ak}[1]{{\color{Orange}{{AK: #1}}}}
\else
 \newcommand{\dhwc}[1]{ }
 \newcommand{\dhws}[1]{ }
 \newcommand{\ak}[1]{ }
\fi

\newcommand{\todoPostArxiv}[1]{}

\makeatother

\newcommand{\ba}{\begin{eqnarray}}
\newcommand{\ea}{\end{eqnarray}}

\newcommand{\eq}[1]{\begin{align}#1\end{align}}

\newcommand{\II}{{\mathcal{I}}}

\newcommand{\LandauerLoss}{{\mathcal{L}}}

\newcommand{\vx}{{\pmb{x}}}

\newcommand{\vP}{{\bold{P}}}

\newcommand{\QQ}{{{\mathcal{Q}}}}

\newcommand{\SSS}{\mathcal{S}}

\newcommand{\pa}{\mathrm{pa}}
\newcommand{\Anc}{\mathrm{Anc}}

\newtheorem{example}{Example}

\begin{document}

\title{Uncertainty relations and fluctuation theorems for Bayes nets
}

\author{David H. Wolpert}
\affiliation{Santa Fe Institute, Santa Fe, New Mexico \\
Complexity Science Hub, Vienna\\
Arizona State University, Tempe, Arizona\\
\texttt{http://davidwolpert.weebly.com}}

\begin{abstract}
Recent research has considered the stochastic thermodynamics of 
multiple interacting systems, representing the overall system as a Bayes net. 
I derive fluctuation theorems governing the entropy production (EP)
of arbitrary sets of the systems in such a Bayes net. I also derive 
``conditional'' fluctuation theorems, governing the distribution of EP 
in one set of systems conditioned on 
the EP of a different set of systems. I then derive
thermodynamic uncertainty relations relating the EP of the overall system 
to the precisions of probability currents within the individual systems.
\end{abstract}

\maketitle

\textit{Introduction.--- } Much of stochastic thermodynamics 
considers a single system executing a specified dynamics, without considering how the system might decompose into
a set of interacting subsystems. Examples include analyses of a single system undergoing bit erasure~\cite{parrondo2015thermodynamics,sagawa2014thermodynamic}, or more generally an arbitrary discrete-time
dynamics~\cite{maroney2009generalizing,wolpert_arxiv_beyond_bit_erasure_2015,owen_number_2018}, as well as
a single system maintaining a non-equilibrium steady state (NESS~\cite{seifert2012stochastic}). In particular, there has
been groundbreaking work resulting in fluctuation theorems (FTs~\cite{rao_esposito_my_book_2019,crooks1998nonequilibrium,jarzynski1997nonequilibrium,esposito.harbola.2007,seifert2012stochastic})
and thermodynamic uncertainty relations (TURs~\cite{liu2019thermodynamic,hasegawa2019generalized,falasco2019unifying,horowitz_gingrich_nature_TURs_2019,gingrich_horowitz_finite_time_TUR_2017,chiuchiu.pigolotti.discrete.time.TURs.2018}) for single systems.

Other research has considered the thermodynamics of two interacting subsystems~\cite{horowitz2014thermodynamics},
in some cases where the first subsystem measures the second one~\cite{horowitz2011designing,sagawa2008second},
or performs a sequence of measurements and manipulations of the second one~\cite{horowitz_vaikuntanathan_PRE_2010,mandal2012work,barato2014stochastic,strasberg2017quantum}. 
In particular, there has been research on FTs for a subystem under the feedback control of another
subsystem~\cite{sagawa2012fluctuation,horowitz_vaikuntanathan_PRE_2010}.

However, many physical systems are most naturally viewed as sets
of more than two interacting subsystems, with their joint discrete-time
dynamics described by a probabilistic graphical model~\cite{koller2009probabilistic}. 
For example, all circuits have this character~\cite{calhoun2008digital,wolpert_thermo_comp_review_2019,wolpert_kolchinsk_first_circuits_published.2020}, including biological 
circuits~\cite{qian2011scaling,yokobayashi2002directed}, and such systems are common in 
biology more generally~\cite{friedman2000using,friedman2004inferring,larjo2013active,lahdesmaki2006relationships,muggleton2005machine,cho2012network,bielza2014bayesian}. The extension 
of stochastic thermodynamics to study such scenarios was  
pioneered in~\cite{ito2013information,ito_information_2015,Otsubo2018}, which modeled the joint discrete-time dynamics of the subsystems
using Bayesian networks (BNs~\cite{koller2009probabilistic,neapolitan2004learning}).
The major result of~\cite{ito2013information} was an FT
governing the entropy production (EP) of any single one of the subsystems in a BN.

In this paper I extend~\cite{ito2013information}
by deriving FTs that govern the aggregate EP of any number of the subsystems in a BN. I also derive ``conditional FTs'' governing
the EP of any set of the subsystems in
a BN, conditioned on a known value of the EP of a separate set of those
subsystems.
%
%
In also derive TURs that relate the total EP generated
by running a BN to the precisions of currents defined separately for each of the subsystems in that 
BN~\cite{barato_seifert_TURs_2015,gingrich_horowitz_finite_time_TUR_2017,horowitz_gingrich_nature_TURs_2019,falasco2019unifying,hasegawa2019generalized}. 



\textit{Stochastic thermodynamics of semi-fixed processes.--- }
The entropy of a distribution $p(X)$ is $S(p(X))$ or $S(p)$~\cite{cover_elements_2012}, and
the entropy at time $t$ is $S_t(X)$. The mutual information of a distribution $p(X, Y)$ is $I_p(X ; Y) = S(p(X)) + S(p(Y)) - S(p(X, Y))$, or just $I(X ; Y)$ for short.
$\delta(.,.)$ is the Kronecker delta, and $|A|$ is the cardinality of set $A$.

The term ``(forward) protocol'' refers to a sequence of Hamiltonians and rate matrices in 
a continuous-time Markov chain (CTMC)~\cite{esposito.harbola.2007,falasco2019unifying,seifert2012stochastic,esposito2010three}. 
The combination of an initial distribution $p^{t_0}(x)$ and a protocol fixes $\vP(\vx)$, the probability density function over trajectories $\vx$ of the system. 
For simplicity I assume there is a single heat bath and choose units so that $k_B T = 1$,
although the extension to multiple reservoirs is straightforward.
%
%
%

A \textbf{semi-fixed} process is any process involving two 
distinct systems, the \textbf{evolving} 
system and the \textbf{fixed} system, with states $x_A$ and $x_B$, respectively, where $x_B$ 
does not change during the interval $[t_0, t_1]$. (Some aspects of such processes are analyzed 
in~\cite{sagawa2010generalized,horowitz2011thermodynamic,sagawa2012fluctuation,wolpert_thermo_comp_review_2019,wolpert2020thermodynamics,Boyd:2018aa}.)
%

$\QQ(\vx)$ is the \textbf{entropy flow} (EF) from the heat bath into the joint system during the
process  if that system follows trajectory $\vx$.
So the \textbf{global EP} 
is~\cite{seifert2012stochastic,van2015ensemble}
\eq{
\sigma(\vx, p^{t_0}, p^{t_1}) &:= \left(  \ln[p^{t_0}(\vx^{t_0})]-\ln[p^{t_1}(\vx^{t_1})]  \right) -\QQ(\vx) 
\label{eq:global_ep_def}
} 
The \textbf{local EP} of the evolving system is
\eq{
{\sigma}_A(\vx, p_A^{t_0}, p_A^{t_1}) &:=
	 \left(\ln[p_A^{t_0}(\vx_A^{t_0})] - \ln[p_A^{t_1}(\vx_A^{t_1})] \right) - \QQ(\vx)
\label{eq:7b}
}
where $p_A$ refers to the marginal distribution over the states of the evolving system.
So in a semi-fixed process,
\eq{
\sigma(\vx, p^{t_0}, p^{t_1}) = {\sigma}_A(\vx_A, p^{t_0}_1, p_A^{t_1})  - \Delta I_{p^{t_0}, p^{t_1}}(\vx_A ; \vx_B)
\label{eq:7a}
}
where $ \Delta I_{p^{t_0}, p^{t_1}}(\vx_A ; \vx_B)$ is the difference between the ending and starting (stochastic) mutual information
between the two systems~\cite{sagawa2012fluctuation,sagawa_generalized_2010}.

A semi-fixed process is a \textbf{solitary process} if the EF only depends on the 
evolving system, i.e., 
\begin{align}
\label{eq:fracsystem3}
\QQ(\vx) =  \QQ_1(\vx_A)
\end{align}
for some function $\QQ_1(\cdot, \cdot)$. The evolving system of a solitary process is a \textbf{solitary} system. 
The local EP of a solitary process is only a function of $\vx_A$.
Concretely, we can assume that the CTMC of a solitary process has the form
\eq{
\dfrac{d p(x_A, x_B)}{dt} &= \sum_{x'_A,x'_B} W_{x_A,x'_A}(t) \delta(x'_B, x_B) p(x'_A, x'_B)
\label{eq:solitary_rate_matrix_form}
}
for some rate matrix $W(t)$.

Standard arguments establish that the expectation $\langle {\sigma}_A \rangle = 
\int d\vx_A \, \vP(\vx_A) {\sigma}_A(\vx_A, p_A)$ is non-negative in a solitary
process~\cite{wolpert_thermo_comp_review_2019,Boyd:2018aa,wolpert2020thermodynamics}.
(This need not be true for more general semi-fixed processes --- see \cref{app:diff_semi_fixed_solitary}.) So by \cref{eq:7a},  the expected
global EP generated by running the joint system is lower-bounded by the expected drop in mutual information, $-\Delta I(X_A ; X_B)
:= -\int d \vx \vP(\vx) \Delta I(\vx_A ; \vx_B)$.
The data-processing inequality~\cite{cover_elements_2012} confirms that this  bound
is non-negative. In fact, typically $-\Delta I(X_1 ; X_2)$ is a strictly positive lower bound on expected global EP
in solitary processes.  

\textit{Stochastic thermodynamics of Bayes nets.--- } 
Suppose we have a system composed of a finite set of separate subsystems
with joint states
$x = (x_1, x_2, \ldots)$. For any times $t, t'>t$, 
the joint distribution at $t'$ is
\eq{
p(x^{t'}_1, x^{t'}_2, \ldots) &= \sum_{x^{t}_1, x^{t}_2, \ldots} p\left(x^{t'}_1, x^{t'}_2, \ldots \bigg| x^{t}_1, x^{t}_2, \ldots \right) p \left(x^{t}_1, x^{t}_2, \ldots \right)  
\nonumber
}
Typically there will be conditional independencies in how each of the subsystems evolve.
In general, this means that we can decompose $p(x^{t'}_1, x^{t'}_2, \ldots | x^{t}_1, x^{t}_2, \ldots) $
into a product of conditional distributions, each of which captures some of 
those conditional
independencies.
As an example, suppose there are three subsystems, $A, B, C$, and that $x_A^{t'}$ is 
statistically independent of $x^t_C$, given the pair of values $x^t_A, x^t_B$. Suppose as
well that $x_C^{t'}$ is 
statistically independent of $x^t_A$, given the pair of values $x^t_C, x^t_B$, and that
$x^{t'}_B = x^t_B$ with probability $1$. Then
\eq{
p\left(x^{t'}_A, x^{t'}_B, x^{t'}_C | x^{t}_A, x^{t}_B, x^t_C\right) = p\left(x^{t'}_A | x^t_A, x^{t}_B\right) p\left(x^{t'}_C | x^{t}_B, x^t_C\right) \delta(x^{t'}_B, x^t_B)  \nonumber
}

\begin{figure}[tbp]
\includegraphics[width=8cm,height=7cm]{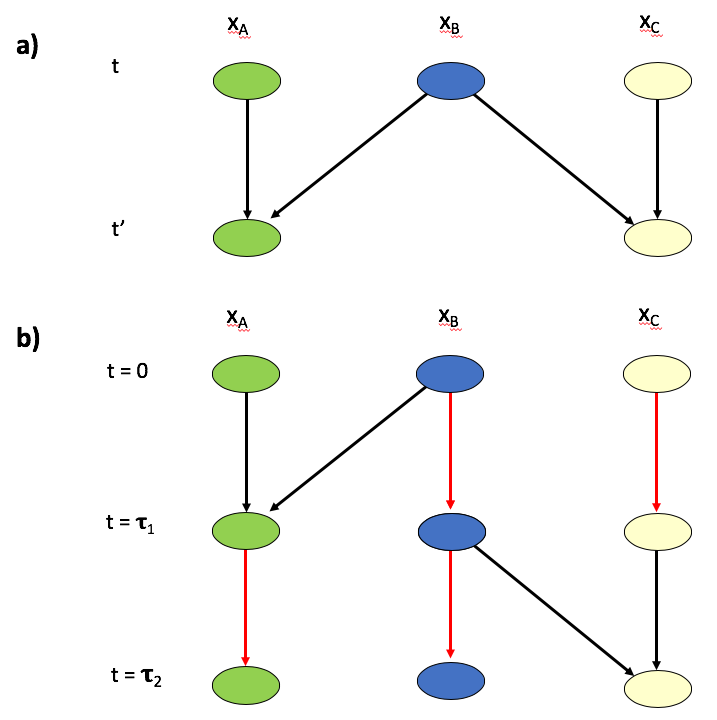}
\caption{a) The example BN discussed in the text. 
b) That BN expanded into a sequence of two solitary processes.
Red arrows indicate the identity map. 
In the first solitary process the evolving system is subsystems $A$ and $B$, while in the 
second one it is $C$ and $B$.}
\label{fig:measure_evolve}
\end{figure}

We can represent this equation
with a directed acyclic graph (DAG), as shown in \cref{fig:measure_evolve}(a). The top three nodes are the \textbf{root nodes} of the DAG,
representing the time-$t$ states of the three subsystems. The bottom nodes are the \textbf{leaf nodes}, representing the time-$t'$
states of two of the subsystems. (Subsystem $B$ does not evolve, so we dispense with
its leaf node.) The directed edges into the bottom-left leaf node indicate that $x^{t'}_A$ depends only on $x^{t}_A$ and $x^{t}_B$,
the two \textbf{parents} of that node. Similarly, the edges into 
the bottom-right leaf node indicate that $x^{t'}_C$ depends only on $x^{t}_C$ and $x^{t}_B$,
the two parents of that node. 
%
%

This representation of a distribution is an example of a {Bayes net}. 
BNs can be generalized to represent the dynamics over an arbitrary number of subsystems. In addition,
they can represent dynamics over an arbitrary number of times, not just the
two times illustrated in \cref{fig:measure_evolve}(a), simply by adding more layers to the DAG. 
This makes them particularly well-suited to modeling the discrete-time thermodynamics of a set of interacting
subsystems, in which a given subsystem may have its state
changed at more than one moment in time. (See \cref{app:BN_details} in the Supplemental Material at [URL will be inserted by publisher] for more
details of BNs, and some technical issues with using them to model the evolution
of physical systems.)

In general, for any BN $G$, there are many equivalent BNs that have the
same conditional distributions at their nodes as $G$, but physically implement
those distributions in a different time order from $G$. Moreover, for any $G$, there is
always such an equivalent BN in which only one node's conditional distribution 
is implemented at a time. (This is called a ``topological order'' of $G$~\cite{koller2009probabilistic}.) An example 
is the BN in \cref{fig:measure_evolve}(b), which implements the same conditional
distributions
as the
BN in \cref{fig:measure_evolve}(a), but in two successive time-intervals rather
than one. 


Keeping with the convention established in~\cite{ito2013information},  I
restrict attention to  BNs where only one node's 
conditional distribution is  implemented at a time.
This means the evolution of any node $v$ in the BN occurs in a solitary process, where  
the evolving system $A$ is the union of the subsystem corresponding to $v$ and the
subsystems corresponding to the parents of $v$. (Note that I refer to the union of $v$ and its parents as
the ``evolving system'' during this solitary process, even though only $v$ changes its state.)
As an example, in \cref{fig:measure_evolve}(b),
the update of subsystem $C$ in the time-interval $[\tau_1, \tau_2]$ is
a solitary process where the evolving system's state is $(x_B, x_C)$, while the state of the fixed system is $x_A$.
(Note from \cref{eq:solitary_rate_matrix_form} that we cannot identify $C$ by itself as the solitary system, since the evolution
of $x_C$ depends on $x_B$ as well as its own state, $x_C$.)
It should be noted that the restriction to sequences of solitary processes is done only to simplify the analysis; the results below can be
extended to allow multiple nodes to update at once, so long as no node is
updated at the same time as one of its parents. (See \cref{app:path_wise_sub} in the Supplemental Material at [URL will be inserted by publisher].)

I write the set of nodes in a BN as $V$, and label them with successive integers.
In general, more than one $v \in V$ might refer to evolution of the same subsystem, just at different times.
In light of the fact that the evolving systems in the solitary processes will be unions of
a node and its parents, for any $V' \subset V$ I define 
$[V'] := V' \cup \pa(V') := \cup_{v\in V'} \; v \cup \pa(v)$, where $\pa(v)$ indicates the parents of node $v$.
In addition, for any $V' \subset V$, I define $-V' := V \setminus V'$.
I write the distribution over all subsystems after the process implementing node $v$ has run
as $p^{{v}}$,
and unless indicated otherwise, assume that it runs in
the time interval $[v-1,v]$. The root nodes are jointly sampled at $t=0$ according to distribution $p^0(x)$, resulting in full trajectory $\vx$.
For any $v \in V$, I
write the segment of $\vx$ corresponding to the time interval when node $v$ runs as $\vx^v$, reserving
subscripts for specification of particular subsystems. 
As an example of this notation, $\vx_{[v]}$ is the full trajectory of the components of $\vx$ specified by $[v]$,
i.e., $\vx_{[v]} \;=\; \vx_{\{v' \in v \cup \pa(v)\}}$, while $\vx^v_{[v]}$ is the segment of that trajectory
in the time interval $[v-1,v]$. 

Let $\pi = \{\pi_v(x_v | x_{\pa(v)})\}$ indicate the set of the conditional distributions at the nodes of the BN.
I write the local EP generated in the associated solitary process as $\sigma_{[v]}$.
Since EP is cumulative over time, by repeated application of \cref{eq:7a}, once for each node in the BN, we see
that the global EP incurred by running all nodes in the BN if the joint system follows trajectory $\vx$ is
\eq{
\sigma(\vx, \pi, p^{0}) &= \sum_{v=1}^{|V|} \bigg[ {\sigma}_{[v]}(\vx_{[v]}^v,\pi_v, p^{{v-1}}) \nonumber \\
		& \qquad\qquad + \left(I_{p^{{v-1}}}\big(x_{[v]}^{v-1};x_{-[v]}^{v-1}\big) 
		-  I_{p^{{v}}}\big(x_{[v]}^v;x_{-[v]}^v\big)\right)  \bigg] \nonumber \\ 
&:= \sum_{v\in V} {\sigma}_{[v]}(\vx^v) - \Delta I^v(\vx)
\label{eq:11aaa}
}
(Note that the superscript $v$ on $\Delta I^v(\vx)$ indicates both the time interval and the evolving system.)

By the data-processing inequality~\cite{cover_elements_2012},  $-\left\langle  \Delta I^v \right \rangle \ge 0$ for all $v$.
Moreover, for each solitary process run at $v$, $\langle {\sigma}_{[v]}\rangle \ge 0$.
 the expected local EP generated by each solitary process is non-negative. 
(This is not 
true for the analogous expected EP considered in~\cite{ito2013information}; see \cref{app:diff_semi_fixed_solitary}
in the Supplemental Material at [URL will be inserted by publisher].) So by \cref{eq:11aaa},
 $-\left\langle \sum_{v\in V} \Delta I^v(\vx) \right \rangle$
is a lower bound on expected global EP. 

Note though that for any specific trajectory $\vx$ and any $v$, 
${\sigma}_{[v]}(\vx^v)$ and/or $-\Delta I^v(\vx)$ may be negative. 
In addition, the sum in  \cref{eq:11aaa} is independent of 
the topological order of the nodes in the BN, so long as 
each distribution $\pi_v$ is fixed as we vary that order. (See 
 \cref{app:topological_order_irrel}
in the Supplemental Material at [URL will be inserted by publisher].)
So $-\left\langle \sum_{v\in V} \Delta I^v(\vx) \right \rangle$ is the same lower bound on the EP no matter what topological order we use.



\textit{FTs for BNs.--- }
As usual, while the FTs below hold in general, to ascribe thermodynamic
meaning to them one needs to assume local detailed balance~\cite{esposito2010threedetailed}.
For simplicity I only present IFTs; DFTs follow in the usual way.
(See \cref{app:fluct_theorems} in the Supplemental Material at [URL will be inserted by publisher] for proofs,
along with some intermediate results.)

To begin, plugging \cref{eq:11aaa} into the usual FTs allows us to relate
 the local EPs when each node runs with the 
associated changes in mutual information
between the states of the evolving and fixed systems:
\eq{
\left\langle \exp \sum_{v \in V} \left(\Delta I^v - {\sigma}_{[v]}\right)  \right\rangle = 1
\label{eq:first_IFT}
}
where the expectation is over all trajectories.
%
\cref{eq:first_IFT} means that the larger the sum of the drops in mutual information when the nodes run, the larger
must be the sum of EPs generated by running those nodes.

In addition, for every node $v \in V$, and for all associated joint values $({\sigma}_{[v]}, \Delta I^v)$ 
that occur with nonzero probability, there is a ``conditional FT'':
\eq{
\left\langle \exp \sum_{v'\ne v} \left( \Delta I^{v'} - {\sigma}_{[v']}\right)
				\right\rangle_{ {\vP}(.| {\sigma}_{[v]}, \Delta I^v)}
		&= 1
\label{eq:37}
}
where the subscript indicates that the expectation is conditional on the given pair of
values, $({\sigma}_{[v]}, \Delta I^v)$. 

\cref{eq:37} concerns the case where 
we are able to measure the EP generated by the evolving system $[v]$ when node $v$ runs, together with the 
associated change in mutual information between the state of the evolving and fixed systems.
It says that if we average over all instances where that EP and mutual information drop have
those known values, then the associated (exponential of the sum of the) EPs
and drops of mutual information when all the \textit{other} nodes besides $v$ are run must average to $1$. 
%
There is also a  set of conditional FTs, which apply if the experimentalist 
can only observe the local EP ${\sigma}_{[v]}$:
\eq{
\left\langle \exp \left(\Delta I^v + \sum_{v'\ne v} \left( \Delta I^{v'} - {\sigma}_{[v']}\right)\right)
				\right\rangle_{ {\vP}(.| {\sigma}_{[v]})}
		&= 1
\label{eq:third_IFT}
}

As an example, suppose we have two bits, $x_A, x_B \in \{-1, 1\}$, each
implemented as a double-welled potential, and we wish
to flip both bits. A bit evolving by itself under a CTMC cannot flip $-1 \leftrightarrow 1$, even approximately~\cite{owen_number_2018,wolpert_spacetime_2019}. However,
we can flip each bit if we have a third binary system, $x_C$,
which can observe and then control each of those bits, one after the other.

In \cref{app:bit_flips} this kind of dual bit-flip is represented as a sequence of two solitary processes,
with evolving systems $(x_A, x_C)$, and then $(x_B, x_C)$. There is no resetting
of any subsystem between the two solitary processes, so the trajectories of the
joint state of the three bits in each of the two processes are statistically coupled.

Assume we know the
initial joint distribution over $(x_A, x_C)$, and the noise levels in the observation
and control steps of the first solitary process. However, do not know how $x_B$ is coupled to
$x_A, x_C$ initially. We also do not know the noise levels during the
second solitary process, in which $x_B$ is flipped. 

%

Assume also that we (the scientist) measure the work done on the system 
during the first solitary process, $W$, and observe $(x_A, x_B, x_C)$ at both the start 
and end of both processes. 
So as discussed in {App.\,F}, we can measure
the local EP during each process, as well as the values of $\Delta I(AC ; B)(\vx) $ and $\Delta I(BC ; A)(\vx)$ (the latter
two being the gains in
stochastic mutual information between the evolving and semi-fixed systems during each process).

Repeatedly run this scenario, collecting all trajectories $\vx$ in which the local
EP in the first process has some particular, arbitrary value, $\sigma^{AC}$. By
\cref{eq:third_IFT}, the average over those $\vx$ of the (exponential of) $\Delta I(AC ; B)(\vx) + \Delta I(BC ; A)(\vx) - \sigma^{BC}(\vx)$
equals $1$ exactly. So for each $\sigma^{AC}$ that occurs with nonzero probability, there is a probabilistic tradeoff
between the sum of the mutual information changes and the local EP of the 
second solitary process, a tradeoff whose details vary depending on $\sigma^{AC}$. 

\cref{eq:first_IFT,eq:37,eq:third_IFT} generalize to FTs concerning sums over subsets of $V$ rather than all of $V$, 
and / or subsets of all the subsystems. Also,
\cref{eq:37,eq:third_IFT} generalize to probability distributions conditioned on simultaneous properties of
multiple nodes $v$, not just one. 

\textit{TURs for BNs.--- }
Each ${\sigma}_{[v]}$ bounds currents generated in the associated system $[v]$,
in accordance with the usual TURs. (This is not 
true for the analogous expected EP considered in~\cite{ito2013information}.) We can exploit this to derive TURs which relate
the EP generated by running a BN to the currents in each evolving system
when it runs. (See
\cref{app:TURs} in the Supplemental Material at [URL will be inserted by publisher] for 
proofs.)

As an illustration, generalize our scenario so 
that the solitary process running each node $v$ takes a total time $\tau_v$  to run, and therefore starts to run at time $t_{v-1} = \sum_{v' \le v-1} \tau_{v'}$.
Let $j_{[v]}^t(\vx)$ be any instantaneous current over $X_{[v]}$ at time $t$.
Write the
time-integrated current over the associated time interval as $J_{[v]}(\vx) = \int_{t_{v-1}}^{t_v} dt  j_{[v]}^t(\vx)$.
Assume that the rate matrix of the full system is constant throughout each of the intervals when a node runs
(although it will change from one such interval to another, in general).
Then we can exploit
a recently derived TUR~\cite{liu2019thermodynamic} to establish that the expected global EP
bounds the precisions of the currents in the subsystems along with
the associated drops in mutual information:
\eq{
\left\langle \sigma \right\rangle
	&\ge   \sum_v\left( \dfrac{2 \tau_v^2 \left\langle  j_{[v]}^{t_v} \right\rangle^2}{{\mbox{Var}}(J_{[v]})} -  
			\left\langle\Delta I^v\right\rangle \right)
\label{eq:26c}
}

This TUR holds without any restrictions on the distributions at the start
or end of the interval when any specific node $v$ runs. However, suppose that
for each node $v$, $p^v = p^{v-1}$, and so in particular
the evolving system's distribution is the same at the start and end of the interval when it evolves.
Also allow the rate matrix when node $v$ runs to vary rather than
stay constant, so long as it is symmetric about the time
$[t^{v-1}, t^v] \, /\,2$. In this situation we can exploit the generalized TUR~\cite{hasegawa2019generalized,falasco2019unifying}
to establish a different TUR for the full BN:
\eq{
\left\langle \sigma \right\rangle 
	&\ge \sum_v 	\left(  \ln \bigg[\dfrac{2 \langle J_{[v]}\rangle^2}{{\mbox{Var}}(J_{[v]})} +1\bigg] - 
			\left\langle\Delta I^v\right\rangle \right)  
\label{eq:26a}
}

Now suppose that in fact every evolving system $[v]$ is in an NESS
when node $v$ runs (though the NESS's may differ depending on which node is running). Then  
\eq{
\left\langle \sigma \right\rangle  
	&\ge   \sum_v\left( \dfrac{2 \langle J_{[v]}\rangle^2}{{\mbox{Var}}(J_{[v]})} -  
		\left\langle\Delta I^v\right\rangle \right)  
\label{eq:26b}
}
This formula holds even if the distribution of the global system is continually changing, 
so long as during each solitary process the marginal distribution of
the associated solitary system does not change.


\label{sec:ex}

\textit{Example of TUR's for a BN.--- } Suppose we have three subsystems, $A, B$ and $C$, 
jointly evolving as in \cref{fig:measure_evolve}(b). So
in the first solitary process the evolving system is the composite system $AB$ and
$C$ is the fixed system, while in the second solitary process the evolving system is the composite system $BC$ and
$A$ is the fixed system. 
Let $J_A(.)$ be any net current among subsystem $A$'s states during the first solitary process, and similarly
for $J_C(.)$, with $j_A^t(.)$ and $j_C^t(.)$ the associated instantaneous currents at time $t$. 
Assume that the rate matrix implementing the first solitary process is constant during that process, and
similarly for the (different) rate matrix implementing the second solitary process. 
So in the first solitary process subsystem $A$ evolves according to a matrix
exponential, where the precise matrix being exponentiated is specified by (the unchanging) value of $x_B$, i.e.,
\eq{
P(x_A^{\tau_1} \,|\, x_A^0, x^0_B) &= \exp{\left(\tau_1 W^{x_B^0}_A\right)}\bigg|_{x_A^{\tau_1}, x_A^{0}}
\label{eq:Aupdate}
}
(and similarly for subsystem $C$ in the second process).

Plugging into \cref{eq:26c} and then using the fact that $x_B$ never changes in either solitary process,
\eq{
\label{eq:15}
\left\langle \sigma \right\rangle 
	&\ge \dfrac{2\tau_1^2 \langle  j_{A}^{\tau_1}\rangle^2}{{\mbox{Var}}(J_A)} 
		+ \dfrac{2 \tau_2^2 \langle  j_{C}^{\tau_1+\tau_2}\rangle^2}{{\mbox{Var}}(J_C)}  -  \left\langle\Delta I^A\right\rangle
				- \left\langle\Delta I^C\right\rangle  \\
	&=  \dfrac{2\tau_1^2 \langle  j_{A}^{\tau_1}\rangle^2}{{\mbox{Var}}(J_A)} 
		+ \dfrac{2 \tau_2^2 \langle  j_{C}^{\tau_1+\tau_2}\rangle^2}{{\mbox{Var}}(J_C)}   \nonumber \\
		&\;\;\;\;+ \Delta S(X_A, X_B, X_C) - \Delta S(X_A, X_B) - \Delta S(X_C, X_B)
\label{eq:62}
}
where here $\Delta$ indicates the value of a quantity at $t=\tau_1 + \tau_2$ minus its value at $t=0$.
\cref{eq:62} provides
a trade-off among global EP, three entropy drops, the instantaneous currents when each of the two subsystems finishes its 
update, and associated net current variances. Note that the RHS of \cref{eq:62} would not change if the two
solitary processes had been run in the reverse order. 

Now suppose $X_B$ is binary, both matrices $W^{x_B}_A$ 
have a (unique) NESS over $X_A$, but that NESS differs for the two $x_B$ values.
Suppose the initial distribution is
\eq{
p^0(x_A, x_B, x_C) &= p^0(x_B) p^0(x_A | x_B) \delta(x_C, x_A) 
}
Next, assume that for both values of $x_B$, $p^0(x_A | x_B)$ is the NESS of the associated rate matrix $W^{x_B}_A$. 
This means that $p^0(x_A, x_B)$ is a NESS during the first solitary process, regardless of $p^0(x_B)$. 
Assume that the second solitary process proceeds the same way, just with subsystem $C$
substituted for subsystem $A$. So we can apply \cref{eq:26b}.

Since there are nonzero probability currents in an NESS, there is nonzero probability that 
the \textit{ending} state of $x_A$ after the first solitary process runs differs from
the state of $x_C$ then. 
However, with probability $1$ their initial states were identical. 
So  there is a drop in (expected) mutual information
between the evolving and fixed systems of the first solitary process. The same is true for the second solitary process. These two drops
in mutual information mean that the {global} system is not in an NESS throughout the full process.
In addition, they provide positive values to two terms on the RHS of \cref{eq:26b}, thereby setting a floor for 
a tradeoff between global EP and the precisions of the two solitary processes.

\textit{Discussion.--- } In this paper I
derive new FTs and TURs 
for systems composed of multiple interacting subsystems. Following~\cite{ito2013information},
I formulate the interactions of those subsystems as a Bayesian network. However, in contrast to~\cite{ito2013information}, 
I identify the evolving systems when each node in the BN is implemented in a way that ensures that the associated EP has the conventional
thermodynamic properties of EP (e.g., that its expectation is non-negative).
This is crucial to the derivation of the FTs and TURs. It also allows me to derive conditional FTs,
involving probabilities of global EP conditioned on a given EP value of one of the subsystems.



\textit{Acknowledgements.--- } I would like to thank Kangqiao Liu, Jan Korbel and Artemy Kolchinsky for helpful discussion. 
This work was supported by the Santa Fe Institute, 
Grant No. CHE-1648973 from the US NSF and Grant No. FQXi-RFP-IPW-1912 from the FQXi foundation.
The opinions expressed in this paper are mine and do not necessarily 
reflect the view of the NSF.

\bibliographystyle{amsplain}
\bibliography{../../../../../LANDAUER.Shared.2016/thermo_refs.main,../../../../../LANDAUER.Shared.2016/thermo_refs_2,../../../../../BIB/refs.main.1.BIB.DIR}

\newpage

\appendix

\section{Technical details of the extension of Bayes nets considered in this paper.}
\label{app:BN_details}

Recall from the main text that we have a physical system comprising a finite set $\mathcal{S}$ of subsystems, 
with joint states $x = (x_1, x_2, \ldots, x_{|\SSS|} \in X$. 
We also have a \textit{separate} BN with a set of nodes $\mathcal{V}$.
I indicate the root nodes of the BN as $R \subset \mathcal{V}$, and write the 
non-root nodes as $V :=  \mathcal{V} \setminus R$. (See~\cite{koller2009probabilistic,neapolitan2004learning} for
formal definitions of BNs and the associated terminology.) 

To connect the BN with $\mathcal{S}$,
we need a function $g : \mathcal{V} \rightarrow \SSS$ that maps each $v \in \mathcal{V}$ to one of the subsystems.
Note that in general $g$ will not be invertible; the same subsystem may change its state in a conditional
distribution specified by more than one node of the BN. Physically, this just means that a given subsystem $i$ may change
its state at more than one time as the process proceeds.
(In the main text, following the convention in~\cite{ito2013information}, $g$ is implicit.)

Write the initial distribution 
of the joint state of the subsystems in terms of the distribution
over the root nodes of the BN, as $p^0(x_{g(R)})$.
So the distribution over $x_{g(v)}$ after any non-root node $v$ runs is
\eq{
&\!\!\!\!\!\!\!\! \sum_{x_{g(R)}} p^0(x_{g(R)}) \; \bigg[ \bigg(\sum_{x_{g(\Anc[v])}}  \prod_{v' \in \Anc[v]} 
			 \pi_{v'} ( x_{g(v')} | x_{g(\pa(v'))} ) \bigg)\nonumber \\
	&\qquad \qquad \qquad \qquad \qquad \qquad \;\; \;\;  \times \pi_{v} ( x_{g(v)} | x_{g(\pa(v))} ) \bigg]
}
where $Anc[v]$ is the set of ancestor nodes of $v$ that are not root nodes.
Below I will often write $x_v$ as shorthand for $x_{g(v)}$.

As described in the text, I follow \cite{ito2013information} and
assume that the conditional distributions of the BN's nodes are implemented one after another, in a sequence 
of solitary processes specified by a
topological order of the BN. I index the nodes by their (integer-valued) position in the topological order, 
with an index $v \in \{1 - |R|,\ldots,|V|\}$, so that the non-root nodes start with $v > 0$.

As an aside, note that one can\textit{not} model the dynamics  
when node $v$ runs as a solitary process where the subsystem $v$ considered by itself is the evolving system. The reason is
that since the dynamics of $x_{g(v)}$ depends on the value of $x_{g(\pa(v))}$ when $v$ runs, \cref{eq:fracsystem3}
does not hold for any function $\QQ_1$ if $\vx_A$ is set to $x_{g(v)}$.

For this framework
to give both a complete and consistent representation of the thermodynamics of a set of evolving subsystems, several
requirements must be met. First, there must be 
exactly one root node $r$ corresponds to each
subsystem, i.e., for all subsystems $i$, $g(r) = i$ for exactly one $r \in R$. Second, 
the joint distribution
$p^0(x_{g(R)})$ must have been sampled before any non-root node $v$ runs. (This ensures that
every subsystem has a well-defined state by the time any non-root node $v$ runs.)

A third, more subtle requirement, reflects the fact that we want every non-root node in the DAG to represent a change in the state of one of the physical systems.
This means that for each root node $r$, corresponding to subsystem $i = g(r)$, all
other nodes $v' \in g^{-1}(i)$ representing states of that subsystem lie on a single path of connected edges leading out of $r$. 
This then implies that for all non-root nodes $v$,
there is one (and only one) parent of $v$ in the DAG, $v'$, such that $g(v') = g(v)$; we interpret the value of
$x_{g(v')}$ when node $v$ starts to run as the initial state of $X_{g(v)}$ at the beginning of a process updating it, 
while the value of $x_{g(v)}$ when node $v$ has finished running is the
state of $X_{g(v)}$ when that update has completed. 


Finally, I require that we can implement the BN in a specific type of topological order: one where 
no subsystem's update at a node $v$ depends on an old state of another subsystem which has been overwritten
by the time that $v$ is executed. Formally, this requirement means that
for all nodes $v$, there is no $v' \in \pa(v)$ and other node $v''$ such that: $g(v') = g(v'')$, $v'$
is an ancestor of $v''$, and $v''$ occurs before $v$ in the topological order. This ensures that the physical
process implementing the BN is Markovian. 

Note that not all BNs can be represented in a topological
order that respects this last requirement. Most simply, suppose we have two subsystems, $X$ and $Y$, with states $x$ and $y$.
Suppose that under the BN governing their joint dynamics, $x^1$ depends on both $x^0$ and $y^0$, 
and $y^1$ also depends on both $x^0$ and $y^0$. So if we
update $X$ first under the topological order, then no physical process
can properly update $Y$ in a subsequent process by coupling to the state of $X$,  since $x^0$ no longer describes the state of $X$.
Similarly, if we update $Y$ first, then no physical process
can properly update $X$ after that by coupling to the state of $Y$  since $y^0$ no longer describes the state of $Y$.
So this requirement is actually a restriction on the BNs being considered in this paper.



These requirements are all assumed in~\cite{ito2013information}, implicitly or otherwise. 
In particular, to see that the analysis in~\cite{ito2013information} also assumes that 
the physical process updating the state of the subsystem $g(v)$ can be treated as a solitary process, where  
the evolving system is $A = g(v) \cup g(pa(v))$, note 
that in Eq.\,4 of~\cite{ito2013information}, the two conditional probabilities on the RHS are 
conditioned only on ${\mathcal{B}}^{k+1}$. So it is being assumed that the entropy flow into the baths
can be written as a function of the state of the physical variable $x$ at the times $k$ and $k+1$, as well as the
(identical) states of the variables corresponding to its parents at those two times, i.e., that condition 2
of a solitary process is being obeyed. 

However, for simplicity I modify the formulation in~\cite{ito2013information}
by allowing the subsystems to have their initial states set in parallel,
by sampling the joint distribution $p^0$ over the root nodes at once, rather than require that those nodes
be sampled independently, one after the other. 
In addition, for convenience I relax the standard BN requirement that the distribution over the root nodes be a product distribution,
whereas~\cite{ito2013information} does not address this issue. Ultimately, allowing
the root nodes to be jointly sampled is just a modeling choice. An alternative, adhering to the conventional BN requirement
that all root nodes be sampled independently,
would be to modify the DA to include a
single extra node that is a shared parent of what (in the framework here) is the set $R$ of root nodes.

The framework defined above is similar to several graphical models in the literature,
including non-stationary dynamic Bayesian networks~\cite{robinson2009non}, time-varying dynamic Bayesian
networks~\cite{song2009time}, and non-homogeneous dynamic Bayesian networks~\cite{dondelinger2013non},
among others.
Nonetheless, for simplicity,
I will simply refer to this structure as a Bayes net, even though that is not exactly accurate.

Finally, as a technical comment, it is important to note that 
there are an infinite set of discrete-time conditional distributions $\pi_v(x_v^{v+1} | x_v^{v})$ that cannot be implemented using
a CTMC over $X_{g([v])}$. Instead, in order to use a CTMC to model the dynamical process that results in that conditional distribution, there must be a set of
extra ``hidden'' states of subsystem $g(v)$, in addition to the ``visible'' states in $X_{g(v)}$, and the CTMC must
couple those two sets of states when it runs that conditional distribution to update that subsystem~\cite{owen_number_2018,wolpert_spacetime_2019}. 

However,
at both $t = v-1$ and $t = v$, the beginning and end of when node $v$ runs,
the state of subsystem $g(v)$ must be visible, i.e., it must lie in $X_{g(v)}$ at those two times. 
(If that weren't the case, then we could not be sure that the discrete time dynamics is actually given by $\pi_v(x_v^{v+1} | x_v^{v})$ operating on $p^v(x^v_{\pa(v)})$.)
Accordingly, any hidden states can be ignored in calculating the drop in mutual information as node $v$ runs. For the same reason, the change in $\ln[p^t(\vx)]$ from
$t = v$ to $t = v+1$ doesn't depend on whether hidden states exist. So the only effect 
of such states on Eq. (5) of the main text
is to modify the EF function of the process updating node $v$, 
$\QQ_v(., .)$ (and similarly for Eq. (1) of the main text.)
Therefore their only effect on Eq.\,(7) in the main text
--- which is the starting point for the results in this paper concerning solitary process 
formulations of BNs ---
is to change the EF function implicitly defining ${\sigma}_{[A]}(., .)$.
Since the detailed form of the EF function is irrelevant for the results in this paper 
in the way that they are stated (the EF function for updating node $v$ is subsumed in the function ${\sigma}_{[v]}(., .)$),
we can ignore hidden states for the purposes of this paper.

\section{Path-wise subsystem processes}
\label{app:path_wise_sub}


In the text, for simplicity, I considered the case where each subsystem updates its state
in a solitary process, during which no other subsystem changes its state. However it is
straightforward to weaken that restriction, to allow multiple subsystems to change their state at
the same time, so long as they are independent of one another during that simultaneous update.
While this flexibility is not exploited in the main text, it is worth describing its thermodynamic properties.
I do that in this appendix.

To begin, write the \textit{multi-information} of a joint distribution over a set of random variables, $p(X_1, X_2, ...)$, as
\eq{
	\II(p) &:= \left[ \sum_i S(p(X_i))\right] - S(p(X)) 
\label{eq:multi_information}
} 
(Multi-information is also sometimes called ``total correlation'' in the literature.)
Mutual information is the special case of multi-information where there are exactly two random variables.
I will use ``path'' and ``trajectory'' interchangeably, to mean a function from time into a state space. In the usual
way, I use the argument list of a probability distribution to indicate what random variables have been marginalized,
e.g., $P(x) = \sum_y p(x, y)$. 

Consider a CTMC governing evolution during time interval $[t_0, t_1]$.
That CTMC is a {\textbf{(path-wise) subsystem process}} if
\begin{enumerate}
\item   The subsystems evolve independently of one another, i.e., the discrete-time conditional distribution over the joint state space is
\begin{align}
\pi(\vx^{t_1} | \vx^{t_0}) = \prod_{i \in \SSS}  \pi_i(\vx_i^{t_1} | \vx_i^{t_0}) 
\label{eq:fracsystem1}
\end{align}
\item There are $|\SSS|$ functions, $\QQ_i(.)$, such that the entropy flow  (EF) into the joint system during the process 
if the full system follows trajectory $\vx$ 
can be written as 
\begin{align}
\label{eq:fracsystem2}
\QQ(\vx) = \sum_{i\in \SSS} \QQ_i(\vx_i)
\end{align}
for all trajectories $\vx$ that have nonzero probability under the protocol for initial distribution $p^{t_0}$.
(Recall that the entire sequence of Hamiltonians and rate matrices is referred to as a ``protocol'' in stochastic
thermodynamics.)
\end{enumerate}
\noindent Intuitively, in a subsystem process the separate subsystems evolve in complete
isolation from one another, with decoupled Hamiltonians and rate matrices.
(See~\cite{wolpert2020thermodynamics,wolpert_thermo_comp_review_2019} for explicit examples of CTMCs 
that implement subsystem processes, for the special case where there are two subsystems.)

I use the term \textbf{(path-wise, subsystem) Landauer loss} to refer to the extra, unavoidable EP 
generated by implementing the protocol due to the fact that we do so with a subsystem process:
\eq{
\LandauerLoss(\vx, p) &:= \sigma(\vx, p) - \sum_{i=1}^N {\sigma}_i(\vx_i,p_{i})  \nonumber \\
	&= \II_{p^{t_0}}(\vx^{t_0} ) -  \II_{p^{t_1}}(\vx^{t_1})  \nonumber \\
	&:= -\Delta \II_p(\vx)
\label{eq:7}
}
where the second line uses condition (2) of path-wise subsystem processes to cancel the EFs. (The reason
for the name is that the ``Landauer cost'' of implementing $\pi(\vx^{t_1} | \vx^{t_0})$ --- the minimal EF needed 
by any physical process that implements that conditional distribution --- is increased
by $\LandauerLoss(\vx, p)$ if we add the requirement that the process obey condition (2) of path-wise subsystem processes.)
Note that even though $\Delta \II_p(.)$ is written as a function of an entire trajectory, its value only depends and
the initial and final states of the trajectory.

Both (expected) subsystem EP and global EP are always non-negative. Moreover, by \cref{eq:7}, 
if  the expected multi-information among the subsystems decreases in a subsystem process, 
the Landauer loss must be strictly positive  --- and so the global
EP has a strictly positive lower bound. This is true no matter how thermodynamically efficiently the individual subsystems evolve.

One  way to understand this phenomenon is to note that in general the Shannon information stored in the initial statistical
coupling among the subsystems will diminish (and maybe disappear entirely) as the process runs.
However, for each subsystem $i$ the rate matrix governing how $x_i$ evolves cannot depend on the states of 
the rest of the subsystems, $x_{-i}$, due to condition (2) of subsystem processes.
So that rate matrix cannot exploit the information in the statistical coupling between the initial states of the subsystems
to reduce the total amount of entropy that is produced as the information about the initial coupling
of the subsystem states disappears. In contrast,
if it were not for condition (2), then the rate matrix governing the dynamics of $x_i$ \textit{could} depend  on the value of $x_{-i}$,
and therefore could exploit that value to reduce the amount of entropy that is produced as the information about the initial coupling
of the subsystem states disappears. (See~\cite{wolpert2020thermodynamics,wolpert_thermo_comp_review_2019}.)

These results do not require that the underlying process generating trajectories
is a CTMC. However, from now on I assume that in fact the dynamics is generated by a CTMC,
so that the conventional fluctuation theorems and uncertainty relations all hold.

\section{The differences between the thermodynamic properties of ${\sigma}_{[A]}$ and ${\sigma}_A$.}
\label{app:diff_semi_fixed_solitary}

When $v$ runs, the set of subsystems $[v] = g(v) \cup g(\pa(v))$ form a solitary system, while the subsystem $g(v)$ is only the 
subsystem that changes its state. This is why I write the local EP generated by $[v]$ when node $v$ runs as ${\sigma}_{[v]}(\vx)$ for short. 
In general, this local EP generated by the set of subsystems $[v]$ when $g(v)$ runs
will differ from the local EP generated by just $g(v)$ when node $v$ runs; they are related by
\eq{
{\sigma}_v(\vx) &= {\sigma}_{[v]}(\vx)+ \Delta^v \ln \left[{p(\vx_{\pa(v)} \,|\, \vx_v)}\right]
\label{eq:independent_EP}
}
where $\Delta^vf[p(\vx)]$ is shorthand for $f[p^v(\vx)] - f[p^{v-1}(\vx)]$.
(See \cref{eq:7b,eq:fracsystem3}.) As described above, the analysis in~\cite{ito2013information} concerns $\sigma_v$ rather than $\sigma_{[v]}$.

The expected value of ${\sigma}_v$ can be negative, in contrast
to the expected value of ${\sigma}_{[v]}$. In addition, while the usual FTs and TURs apply
to the EP ${\sigma}_{[v]}$, they do not apply to ${\sigma}_v$ in general.
This is why I formulate the results in the text in terms of ${\sigma}_{[v]}$. However, if desired these results can be recast
in terms of ${\sigma}_v$, by using \cref{eq:independent_EP}.
This allows the results of this paper to be connected with those
in~\cite{ito2013information}.

In the rest of this appendix I discuss this relationship between the two kinds of EP
in more detail.
The first thing to note is that there are several specific thermodynamic properties of the EP of the evolving system ${\sigma}_{[A]}$ in a solitary
process that need not hold for the EP of the evolving system ${\sigma}_A$ in a general semi-fixed process.
Perhaps the most important is that 
in keeping with the conventional second law, in a solitary process the expected subsystem EP
of the evolving system is non-negative. i.e., $\langle {\sigma}_A \rangle \ge 0$, whereas
in an arbitrary semi-fixed process $\langle {\sigma}_A \rangle$ can be strictly negative. 

As an explicit demonstration of such a case where the expected EP can be negative, suppose that the entire joint system 
evolves in a thermodynamically
reversible process, so that $\langle \sigma\rangle = 0$. (Note that this is not possible in general for a solitary process.) Then $\QQ(\vx, p^{t_0}) = \ln[p^{t_0}(\vx)]
- \ln[p^{t_1}(\vx)]$. Therefore 
\eq{
\langle {\sigma}_A \rangle &= S_{{t_1}}(X_A) - S_{{t_0}}(X_A) - \left[S_{{t_1}}(X_A,X_B) - S_{{t_0}}(X_A,X_B)\right]
		\nonumber \\
	&= S_{{t_0}}(X_A | X_B) - S_{{t_1}}(X_A | X_B) 
}
\textit{A priori}, this drop in the conditional entropy of the evolving system's state given the
fixed system's state can be positive or negative.

A second important difference arises if we consider
the minimal amount of work required to send the ending joint
distribution of a semi-fixed process, $p_f$, back to the initial one, $p_i$.
The difference between the amount of work expended in getting from $p_i$ to $p_f$ in
the first place and this minimal amount of work to go back is sometimes called the ``dissipated'' work in going
from $p_i$ to $p_f$, 
because it is the minimal amount of work lost to the heat bath if one were to run a full cycle $p_i \rightarrow p_f \rightarrow p_i$.
Much of the stochastic thermodynamics literature presumes that dissipated work can be treated as a synonym for EP.
In agreement with this, the dissipated work in a solitary process always equals the expected subsystem EP, $\langle{\sigma}_{[A]} \rangle$.
In contrast, dissipated work does not equal $\langle\overline{\sigma}_i\rangle$ in semi-fixed processes, in general. 

Another difference is that in a solitary process
the conventional FT holds for the evolving system with state space $X_A$, considered in isolation from the fixed system,
if the EP in that FT is identified as ${\sigma}_{[A]}(\vx_1, p_1)$. However, in general the conventional FT
will not hold in general for a semi-fixed process with state space $X_A$
in an arbitrary semi-fixed process, if the EP in that FT is identified as ${\sigma}_A(\vx, p)$.

As a final example, the expected EP of the evolving system in a solitary process bounds the precision of any current defined over
the state of the subsystem, in the usual way given by the thermodynamic uncertainty relations~\cite{falasco2019unifying}.
However, the expected EP of the evolving system in a semi-fixed process need not have so simple
a relationship with the current in that subsystem.


\begin{example}
\label{ex:1}
Consider a process involving three subsystems, $A$, $B$ and $C$.
Only subsystem $A$ changes its state in this process, and the dynamics of subsystem $A$ depends on
the state of subsystem $B$, but \textit{not} on the state of subsystem $C$. We can 
formulate this process as a semi-fixed process
where either subsystem $A$ or the joint subsystem $AB$ is the evolving system.

Note though that we can\emph{not} 
identify subsystem $A$ as the evolving system of a solitary process, with the joint subsystem $BC$ being the fixed 
subsystem. (Since the evolution of subsystem $A$ depends on the state of subsystem
$B$, condition (1) would be violated.) 

On the other hand, the situation is not so clear-cut if we ask whether the process
is a solitary process with $AB$ the evolving system. If
the joint subsystem $AB$ is physically decoupled from subsystem $C$, with no interaction Hamiltonian coupling
$C$ to the other subsystems,
and no coupling of $C$ with $AB$ in the rate matrix for the full system $ABC$, then we have a solitary process, with
the evolving system identified as the joint subsystem $AB$. However, as an alternative, we
could run the entire process in a way that is globally thermodynamically reversible, incurring zero global EP.
(N.b., in general this would require an interaction Hamiltonian coupling
$C$ to the other subsystems,
and require that the rate matrix for the full system $ABC$ couples the dynamics of $C$ to that of $AB$.)
In this case, the bound in \cref{eq:7a} would be violated in general if we identify $AB$ as
the evolving system (e.g., if the expected drop in mutual information is strictly
nonzero). So the overall process would not a solitary process with $AB$ the evolving system.

This demonstrates that in general, just specifying the joint  dynamics of a co-evolving set of subsystems does 
not determine whether we can
view a particular physical process that implements that dynamics as a solitary process, for some appropriately
identified evolving system.
\end{example}

\section{Global EP is independent of the topological order of the nodes in the BN}
\label{app:topological_order_irrel}

By hypothesis, for all $v \in V$, the physical process that implements
the conditional distribution $\pi_v(x_{g(v)} | x_{g(\pa(v))})$ is the same in any two
topological orders. So changing the topological order doesn't change any
of the values ${\sigma}_{[v]}$. Therefore to establish the claim, we need to establish
that $-\sum_{v\in V} \Delta I^v(\vx)$ is independent of the topological order.

To do that, given a topological order, label the nodes $v \in V$ in the sequence they occur
in that topological order as $1, 2, \ldots |V|$, so that they occur in corresponding
time intervals $[0, 1], [1, 2], \ldots, [|V| - 1, |V|]$. (Note that in general it may be that
more than one of those nodes change the state of the same subsystem.) Express $g(.)$ accordingly.
Introduce the shorthand that $g([v]) = \cup_{v' \in [v]} g(v')$. So it is the union of all the subsystems
that are in the evolving system when the subsystem $g(v)$ changes its state.

Next, use the fact that while
the marginal entropy of the evolving system changes during a solitary process, the
marginal entropy of the semi-fixed system doesn't change, to expand
\begin{widetext}
\eq{
-\sum_{v\in V} \Delta I^v(\vx) \;=\; &-\left[\left(\ln(p^0_{g([1])}(\vx)) + \ln(p^0_{-g([1])}(\vx)) - \ln(p^0(\vx))\right)
									- \left(\ln(p^1_{g([1])}(\vx)) + \ln(p^0_{-g([1])}(\vx)) - \ln(p^1(\vx)\right)\right] \nonumber \\
			 &\qquad - \left[\left(\ln(p^1_{g([2])}(\vx)) + \ln(p^1_{-g([2])}(\vx)) - \ln(p^1(\vx))\right)
									- \left(\ln(p^2_{g([2])}(\vx)) + \ln(p^1_{-g([2])}(\vx)) - \ln(p^2(\vx))\right)\right] \nonumber \\
			 &\qquad - \ldots \nonumber \\
			 &\qquad - \left[\left(\ln(p^{|V|-1}_{g([|V|])}(\vx)) + \ln(p^{|V|-1}_{-g([|V|])}(\vx)) - \ln(p^{|V|-1}(\vx))\right)
									- \left(\ln(p^{|V|}_{g([|V|])}(\vx)) + \ln(p^{|V|-1}_{-g([|V|])}(\vx)) - \ln(p^{|V|}(\vx))\right)\right] \nonumber \\
	\;=\; & \ln(p^0(\vx)) - \ln(p^{|V|}(\vx)) 	- \sum_v \ln(p^{v-1}_{g([v])}(\vx)) - \ln(p^{v}_{g([v])}(\vx)) 
\nonumber \\
\label{eq:topo_order_1}
}
\end{widetext}
where the sums in last two equations
are over the set of subsystems, and for simplicity I assume that each subsystem $i$ occurs in at least one node, i.e., 
for all subsystems $i$, $g^{-1}(i) \ne \varnothing$.

The RHS of \cref{eq:topo_order_1} is fully specified by the combination of the BN and the initial distribution.
So it cannot depend on the topological order. This establishes the claim.


\section{Derivations of fluctuation theorems for Bayesian networks}
\label{app:fluct_theorems}

Let $\tilde{\vx}$ indicate the time-reversal of the trajectory $\vx$. (For simplicity, I restrict attention to spaces $X_i$
whose elements are invariant under time-reversal.) Let $\vP(\vx)$
indicate the probability (density) of $\vx$ under the forward protocol running the entire BN. Let $\tilde{\vP}(\vx)$ indicate the probability of the same trajectory if we run the protocol in time-reversed order, where the ending distribution over $X$ under
$\vP$ is the same as the starting distribution under $\tilde{\vP}$. Also write $\tilde{\vx}^v$ to indicate the time-reversal
of the trajectory segment $\vx^v$.

In the next subsection, I derive fluctuation theorems concerning probabilities of trajectories, and
in the following subsection, I derive fluctuation theorems concerning the joint probability that
each of the subsystem EPs has some associated specified value.

\subsection{Fluctuation theorems for trajectories}

Plugging Eq.\,(7) in the main text
into the usual detailed fluctuation theorem (DFT)~\cite{van2015ensemble}
gives the DFT,
\eq{
 \ln \bigg[  \dfrac{ \vP(\vx)} { \tilde{\vP}(\tilde{\vx})}  \bigg] 		
	 	&= 	
		 \sum_{v=1}^{|V|} {\sigma}_{[v]}(\vx^v) - \Delta I^v({\vx})
\label{eq:true_18}
}
for all $\vx$ with nonzero probability under $\vP$.
Exponentiating both sides of \cref{eq:true_18} and then integrating
results in the integrated fluctuation theorem (IFT),
\eq{
\left\langle e^{ - \sum_{v=1}^{|V|} {\sigma}_{[v]} - \Delta I^v} \right\rangle 	&:= 
	\int d\vx \, \vP(\vx)   e^{  \left(-\sum_{v=1}^{|V|} {\sigma}_{[v]}(\vx^v) - \Delta I^v(\vx)\right)}
				\nonumber \\
&= 1
\label{eq:19a}
}

In addition to applying to runs of the entire BN, 
the usual DFT applies separately to successive time intervals, 
i.e., to each successive interval at which exactly one node and its parents
co-evolve as that node's conditional distribution is executed. Therefore for all $v$,
\eq{
 \ln \bigg[  \dfrac{ \vP(\vx^v)} { \tilde{\vP}(\tilde{\vx}^v)}  \bigg]
		&= {\sigma}_{[v]}(\vx^v) - \Delta I^v(\vx)  
\label{eq;19a}
}
which results in an IFT analogous to \cref{eq:19a}. 


Combining \cref{eq:true_18,eq;19a} gives
\eq{
\II(\vP(\vx)) &= \II(\tilde{ \boldsymbol{P}}(\tilde{\vx})) 
\label{eq:4}
}
where I define
\eq{
\II(\vP(\vx)) &:= \ln \left[\dfrac{\vP(\vx)}{\prod_{v=1}^{|V|} \vP(\vx^v)} \right]
\label{eq:17c}
}
and similarly for $\II(\tilde{\vP}(\tilde{\vx}))$. 
Note that \cref{eq:4} can be rewritten as
\eq{
\frac{\vP \left( \vx \right )}
	{ \tilde{\vP} \left( \tilde{\vx} \right)}
	&= \prod_v \frac{\vP \left(\vx^v \right)} {\tilde{\vP} \left(\tilde{\vx}^v \right)}
\label{eq:19bb}
}
(This equality can also be derived directly, without invoking DFTs~\footnote{To see this, first expand
$\vP(\vx) = \vP(\vx(0)) P(\vx^1 | \vx(0))  P(\vx^2 | \vx(1)) \ldots$ where $\vx^k$ means the time interval  $[k -1, k]$ and 
and $\vx^t$ means the slice of $\vx$ at time $t$. 
Expand $\tilde{\vP}(\tilde{\vx})$ similarly. Then
note that by construction, $\vP(\vx(t)) = \tilde{\vP}(\tilde{\vx}(|V| - t))$ for all $t$. Combine
these facts to expand the ratio $\vP(\vx) / {\tilde{\vP}}(\tilde{ \vx})$, and then take take its logarithm.
}.)

The quantity $\II(\vP(\vx))$ defined in \cref{eq:17c}
is an extension of multi-information to concern
probabilities of {entire trajectories} of the joint system. So loosely speaking, \cref{eq:4} means that the amount of
information that the set of all the trajectory segments $\{\vx^v\}$ provide about one another (under $\vP$)
equals the amount of information that the set of all the trajectory segments $\{\tilde{\vx}^v\}$ provide about one another
(under $\tilde{\vP}$). (Note that the same subsystem may evolve in more than trajectory segment. )
In this sense, there is no arrow of time, as far as probabilities of trajectory segments are concerned.

We can also combine  \cref{eq:true_18,eq;19a} to derive
DFTs and IFTs involving conditional probabilities, in which
the trajectories of one or more of the subsystems are given. To illustrate this,
pick any $V' \subset V$, and plug \cref{eq;19a} into the sum on the RHS
of \cref{eq:true_18} for all of the nodes $v \in V'$. Define $\vx^{V'} := \{\vx^{v'} : v' \in V'\}$,
i.e., the ``partial trajectory'' given by all segments $v' \in V'$ of the trajectory $\vx$.
Then after clearing terms and using \cref{eq:19bb},
we get the following {conditional DFT}, 
which must hold for all partial trajectories $\vx^{V'}$ with nonzero probability under
$\vP$:
\eq{
 \ln \left[  \dfrac{ \vP\left(\vx | \vx^{V'}\right)} { \tilde{\vP}\left(\tilde{\vx} | 	
		\tilde{\vx}^{V'}\right)}  \right]   &=
			\II(\tilde{\vP}(\tilde{\vx}^{V'})) - \II(\vP(\vx^{V'})) +
		 \sum_{v \in V \setminus V'} {\sigma}_{[v]}(\vx^v) - \Delta I^v(\vx) 
\label{eq:16}
}
In turn,
\cref{eq:16} gives the following conditional IFT which must hold for each partial
trajectory $\vx^{V'}$ with nonzero probability under $\vP$:
\eq{
 \left\langle e^{\left(  \II^{V'} - \tilde{\II}^{V'} + 
			\sum_{v \in V \setminus V'} \left(\Delta I_{v} - {\sigma}_{[v]}\right) \right)} \right\rangle_{\vP(.|\vx^{V'})}
			&= 1
\label{eq:17}
}
where $\II^{V'}$ is shorthand for the random variable $\II(\vP(\vx^{V'}))$ and similarly for $ \tilde{\II}^{V'}$.
(Note that all terms in the exponent in \cref{eq:17} are defined in terms of the full
joint distributions $\vP(\vx)$, but that $\vx$ is averaged according to $\vP(\vx|\vx^{V'})$.)

Note that in addition to these results which hold when considering the entire system $X$, since each subsystem evolves in a solitary process,
the usual DFT and IFT must hold for each subsystem $v$ considered in isolation, in the interval 
during which it runs. 
So for example,
\eq{
 \ln \bigg[  \dfrac{ \vP\left(\vx^v_{[v]}\right)} { \tilde{\vP}\left(\tilde{\vx}^v_{[v]}\right)}  \bigg] 		&= 	
		{\sigma}_{[v]}(\vx^v)
\label{eq:19b}
}
(Compare to \cref{eq;19a}.) \cref{eq:19b} gives us an additional set of conditional DFTs and IFTs. For example, it gives 
the following variant of \cref{eq:16}
\eq{
\ln \bigg[  \dfrac{ \vP\left(\vx
| \vx^v_{[v]}\right)} 
			{ \tilde{\vP}\left(\tilde{\vx}
| \tilde{\vx}^v_{[v]}\right)}  \bigg]
	&= - \Delta I_{v}(\vx) + \sum_{v' \ne v} \left({\sigma}_{[v']}(\vx^{v'}) - \Delta I_{v'}(\vx)\right)
\label{eq:19c}
}
and the following variant of \cref{eq:17}
\eq{
\left\langle e^{-I_v +
\sum_{v'\ne v} \left( \Delta I_{v'} - {\sigma}_{[v']} \right) } 
				\right\rangle_{\vP(.|\vx^v_{[v]})}
		&= 1
\label{eq:19cc}
}

Note that the numerator of the expression inside the logarithm on the LHS of \cref{eq:19c} is a distribution
conditioned on 
the joint trajectory of (the subsystem corresponding to) node $v$ and its parents when node $v$ runs.
In contrast, the numerator inside the logarithm on the LHS of  \cref{eq:16} is a distribution
conditioned on 
the joint trajectory of \textit{all} of the subsystems when node $v$ runs (not just the joint trajectory of $v$ and its parents).


\subsection{Fluctuation theorems for EP}

We can use the DFTs of the previous subsection, which concern probabilities of trajectories, to construct 
``joint EP DFTs'', which instead concern probabilities of vectors of the joint amounts of EP generated by all of the subsystems. (See Sec.\,6 in~\cite{van2015ensemble}).

To begin, define $\tilde{\Delta} I^v (\tilde{\vx}) := - \Delta I^v(\vx)$. Similarly define
\eq{\tilde{\sigma}_{[v']}(\tilde{\vx}) &:= \ln \left[\dfrac{\tilde{\vP}(\tilde{\vx}^v_{[v]}) } {\vP(\vx^v_{[v]}) }\right]
}
In the special case that $p^{{v}} = p^{{v-1}}$, we can rewrite this as
${\sigma}_{[v]}(\tilde{\vx}_{[v]}, \tilde{\pi}_v, p^{{v}})$, 
the EP generated by running (the part of the protocol that implements) the conditional distribution
at node $v$ backwards in time, starting from the distribution over $X_{[v]}$
that is the \textit{ending} distribution when node $v$ is implemented going forward in time. We cannot
rewrite it that way in general though; see discussion of Eq.\,85 in~\cite{van2015ensemble}.

%

Using this notation, we can now parallel Eq.\,(83) in~\cite{van2015ensemble}. 
By \cref{eq:true_18}, \cref{eq:19b}, and then the fact that the Jacobian
for the transformation from $\vx$ to $\tilde{\vx}$ equals $1$, 
for any set of real numbers $\{\alpha_v, \gamma_v : v \in V\}$, 

\begin{widetext}
\eq{
\vP \left( {\vx} :  
		\left\{ {\sigma}_{[v]}(\vx_{[v]}) = \alpha_v, \Delta I^v(\vx) = \gamma_v : v \in V \right\} \right) 
	&= \int d\vx \boldsymbol{P}(\vx) 
			\prod_{v} \delta\left(  {\sigma}_{[v]}(\vx^v_{[v]}) - \alpha_v \right)
					\delta\left(  \Delta I^v(\vx) - \gamma_v \right)	\nonumber \\
	&= e^{\sum_v \alpha_v - \gamma_v} \int d\vx \tilde{\vP}(\tilde{\vx}) 	\prod_{v} \delta\left(  {\sigma}_{[v]}(\vx^v_{[v]}) - \alpha_v \right)
					\delta\left(  \Delta I^v(\vx) - \gamma_v \right)	\nonumber \\
	&= e^{\sum_v \alpha_v - \gamma_v} \int d\vx \tilde{\vP}(\tilde{\vx}) 	
		\prod_{v} 	\delta\left( \ln \left[\dfrac{\vP(\vx^v_{[v]}) } {\tilde{\vP}(\tilde{\vx}^v_{[v]}) }\right] - \alpha_v \right)
			\delta\left(  \Delta I^v(\vx) - \gamma_v \right) \\
	&= e^{\sum_v \alpha_v - \gamma_v} \int d\tilde{\vx} \tilde{\vP}(\tilde{\vx}) 	
		\prod_{v} 	\delta\left( \ln \left[\dfrac{\tilde{\vP}(\tilde{\vx}^v_{[v]}) } {\vP(\vx^v_{[v]}) }\right] + \alpha_v \right)
					\delta\left( \tilde{\Delta} I^v(\tilde{\vx}) + \gamma_v \right)
												\nonumber \\
	&= e^{\sum_v \alpha_v - \gamma_v} \tilde{\vP} \left(
		\left\{ \tilde{\sigma}_{[v]}(\tilde{ \vx}_{[v]}) = -\alpha_v, \tilde{\Delta} I^v(\tilde{\vx}) = -\gamma_v : v \in V \right\} \right)
\label{eq:big}
}
where the Dirac delta functions involving $\alpha_v$ and $\gamma_v$ in the integrand can be defined as shorthand, e.g., for
derivatives with respect to $\alpha_v$ and $\gamma_v$ of a modification of the integral, in which the associated Dirac delta functions are replaced by
Heaviside step functions.
\end{widetext}
We can write \cref{eq:big} more succinctly as
\eq{
\ln \left[\frac{ {P} \left(\left\{ {\sigma}_{[v]} = \alpha_v,  \Delta I^v = \gamma_v : v \in V \right\} \right)}
	{ \tilde{ {P}} \left( \left\{ \tilde{\sigma}_{[v]} = -\alpha_v, \tilde{\Delta} I^v = -\gamma_v : v \in V \right\} \right)}\right] 
	&= {\sum_{v'} \alpha_{v'} - \gamma_{v'}} 
}
or just
\eq{
\ln \left[\frac{ {P} \left( \left\{ {\sigma}_{[v]}, \Delta I^v \right\} \right)}
	{ \tilde{ {P}} \left( \left\{ -\tilde{\sigma}_{[v]}, -\tilde{\Delta} I^v  \right\} \right)} \right] 
	&= {\sum_{v} \left({\sigma}_{[v]} - \Delta I^{v}\right)} 
\label{eq:30}
}
for short. (Since the arguments of the probabilities in these equations are not full trajectories,
I am indicating those probabilities with $P$ rather than the density function $\vP$.) This confirms 
Eq. (7)
of the main text:
\eq{
\langle e^{\sum_v \Delta I^v - {\sigma}_{[v]}} \rangle := 1
}

In addition to \cref{eq:30}, which concerns the entire BN, 
the conventional extension of the DFT must hold separately for the time interval when each evolving system 
$[v] = \pa(v) \cup \{v\}$ runs:
\eq{
\ln \left[\dfrac{ {P} \left({\sigma}_{[v]}, \Delta {I}_v \right)}
	{{\tilde {P}} \left(-\tilde{\sigma}_{[v]}, -\tilde{\Delta} I^v \right)}\right] 
							&= {\sigma}_{[v]} - \Delta {I}^v
\label{eq:19}
} 
Combining \cref{eq:30,eq:19} establishes that
\eq{
\frac{ {P} \left( \left\{ {\sigma}_{[v]}, \Delta I^v \right\} \right)}
	{ \tilde{ {P}} \left( \left\{ -\tilde{\sigma}_{[v]}, -\tilde{\Delta} I^v  \right\} \right)}
	&= \prod_v \frac{ {P} \left({\sigma}_{[v]}, \Delta I^v \right)} {\tilde{ {P}} \left( -\tilde{\sigma}_{[v]}, -\tilde{\Delta} I^v \right)}
\label{eq:29}
}
(Note that it is \textit{not} true that
$
 {P} \left( \left\{ {\sigma}_{[v]}, \Delta I^v \right\} \right)
	= \prod_v  {P} \left({\sigma}_{[v]}, \Delta I^v \right)
$ in general.) \cref{eq:29} should be compared to \cref{eq:19b}. 

Taking logarithms of both sides of \cref{eq:29} gives 
\eq{
\II( {P}(\{{\sigma}_{[v]}, \Delta I^v\})) &= \II(\tilde{ {P}}(\{-\tilde{\sigma}_{[v]}, -\tilde{\Delta} I^v\})) 
\label{eq:29a}
}
Loosely speaking, \cref{eq:29a} equates two ``amounts of information''. One is the amount of
information that the set of all pairs, \{EP generated by running a particular subsystem, associated drop in mutual information
between that subsystem's state and all other variables in the full system\}
provide about one another. The other is the amount of
information that the set of all pairs, \{EP generated by running a particular subsystem time-reversed, associated gain in mutual information
between that subsystem's state and all other variables in the full system\}
provide about one another.

Combining \cref{eq:30,eq:19} also
gives a set of conditional fluctuation theorems, analogous to \cref{eq:16,eq:17}, only conditioning
on values of EP and drops in mutual information rather than on components of a trajectory. For example,
subtracting \cref{eq:19} from \cref{eq:30} gives the conditional DFT,
\eq{
\!\!\!\ln \left[\frac{ {P} \left( \left\{ \sigma_{[v']}, \Delta I^{v'} : v' \ne v \right\} \;\bigg|\; {\sigma}_{[v]}, \Delta I^{v} \right)}
	{ \tilde{ {P}} \left( \left\{ -\tilde{\sigma}_{[v']}, -\tilde{\Delta} I^{v'} : v' \ne v \right\} 
								\;\bigg|\; -\tilde{\sigma}_{[v]}, -\tilde{\Delta} I^{v} \right)} \right] 
		&= \sum_{v' \ne v} ({\sigma}_{[v']} - \Delta I^{v'}) 
\label{eq:23}
}
which must hold for all quadruples $({\sigma}_{[v]}, \Delta I^v, \tilde{\sigma}_{[v]}, \tilde\Delta I^v)$ that have nonzero probability under $ {P}$. This in turn 
establishes \cref{eq:37} in the main text:
\eq{
\left\langle e^{
\sum_{v'\ne v} \left( \Delta I^{v'} - {\sigma}_{[v']}\right)} 
				\right\rangle_{ {P}(.| {\sigma}_{[v]}, \Delta I^v)}
		&= 1
\label{eq:E25}
}
which must hold for all $({\sigma}_{[v]}, \Delta I^v)$ with nonzero probability under $ {P}$.

As usual, since each subsystem evolves in a solitary process,
the usual DFTs and IFTs must hold for each subsystem $v$ considered in isolation, in the interval 
during which it runs. So for example, 
\eq{
\ln \left[\dfrac{ {P} \left({\sigma}_{[v]} \right)}
	{{\tilde {P}} \left(-\tilde{\sigma}_{[v]} \right)}\right] 
							&= {\sigma}_{[v]} 
\label{eq:19bbb}
}
(Compare to \cref{eq:19}.)  Combining \cref{eq:30,eq:19bbb} gives us an additional set of DFTs and IFTs. For example, it gives 
the following variant of \cref{eq:23}: 
\eq{
\ln \left[\frac{ {P} \left( \left\{ \sigma_{[v']}, \Delta I^{v'} : v' \ne v \right\} \;\bigg|\; {\sigma}_{[v]} \right)}
	{ \tilde{ {P}} \left( \left\{ -\tilde{\sigma}_{[v']}, -\tilde{\Delta} I^{v'} : v' \ne v \right\} 
								\;\bigg|\; -\tilde{\sigma}_{[v]}\right)} \right] 
		&=  - \Delta I^{v} +   \sum_{v' \ne v} ({\sigma}_{[v']} - \Delta I^{v'}) 
\label{eq:23bbb}
}
This immediately gives \cref{eq:third_IFT} in the main text.

%

\section{Analysis of dual bit-flips fluctuations}
\label{app:bit_flips}

To model the dual bit flip scenario in more detail, taking into account
inevitable noise, write $p_i^t$ as shorthand for the  vector giving
the probability of subsystem $i$ at time $t$, and similarly if $i$ is more than
one subsystem. Suppose that in timestep $1$, lasting from $t=0$ to $t=1$, $C$ observes $A$ in
some noisy process, with small (but nonzero) random back-action on $A$. So 
the joint distribution over $x_A, x_C$ gets updated in this time step by a stochastic matrix $\pi$:
\eq{
p_{AC}^1 = \pi p_{AC}^0
}
The more accurate the observation process is, the closer $\pi$ is to a delta function,
setting $x_C = x_A$ with probability $1$, while leaving $x_A$ unchanged.

Then in time step $2$, lasting from $t=1$ to $t=2$, another biased process flips $x_A$. I write the conditional distribution
of this process as $\rho$:
\eq{
p_{AC}^2 = \rho p_{AC}^1
}
The less noise there is in this process, the greater the probability under $\rho$ that $x_A^2 = -x_C^1$.
Assume that the dynamics of $B$ during these two time steps is independent of the states of
$A$ and $C$, though otherwise unknown.

\begin{figure}[tbp]
\includegraphics[width=8cm,height=7cm]{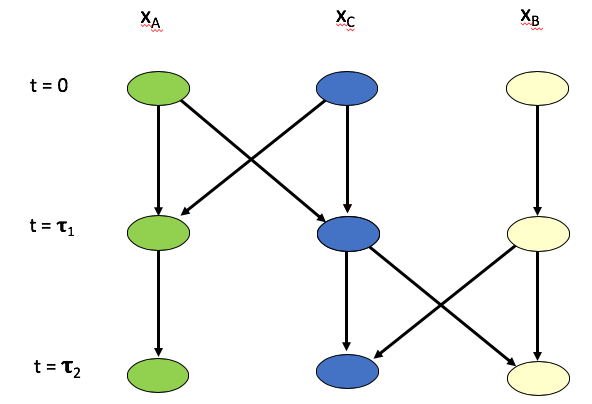}
\caption{The dual bit flip scenario represented as a Bayes net. The first two
solitary processes, in which $C$ couples with $A$ to flip $A$'s state, are grouped together in the BN,
represented by the transition from the $t=0$, root nodes, to their children,
the $t=\tau_1$ nodes. The third and fourth solitary processes, which $C$ couples with $B$ to flip $B$'s state, are grouped together in the BN,
represented by the transition from the $t=\tau_1$ nodes, to their children,
the $t=\tau_2$ nodes.}
\label{fig:bit_flips}
\end{figure}

In time steps $3$ and $4$, the processes of time steps $1$ and $2$ are repeated, perhaps
using stochastic matrices different from $\pi$ and $\rho$, just this time system $C$ 
observes and acts on bit $x_B$, not $x_A$. Note that in the first two time steps, $x_A, x_C$ evolves
as a solitary process, while in the second two time steps, $x_B, x_C$ does. The Bayes net
for this sequence of two solitary processes is shown in \cref{fig:bit_flips}. 

Assume there are two constants $E, E'$ such that at the ends of all time steps, all energy wells have 
depth $E$, and all energy barriers have heights $E'$. By conservation of energy,
this means that the total heat flow from the bath during the first solitary process is the negative of the
work done  on the system during that process. 

Assume that we know the initial joint distribution $p^0_{AC}$ and the matrices $\pi$ and $\rho$, but no other
distributions. In addition, to this dynamic information, we can measure the work in the first solitary process, 
as well as the joint values of $x_A$ and $x_C$ at the beginning and end of that process. We cannot
measure the work in the second solitary process, nor the states of any subsystems at any times during
that process (other than the state $x_{AC}^2$). In addition, we can never measure $x_B$.

For any trajectory $\vx$, by our hypothesis that we can measure $x^0_{AC}$ and  $x^1_{AC}$
and can measure the work done on the system in that trajectory,
we can measure the local EP of the first solitary process:  
\eq{
\sigma^{AC} = \ln\left[p^0_{AC} (x^0_{AC}) \right] - \ln\left[ [\rho \pi p^0_{AC}] (x^1_{AC}) \right] + W(\vx)
} 
(where $[\rho \pi p^0_{AC}] (x^1_{AC})$ means the vector $\rho \pi p^0_{AC}$ evaluated for
the component $x^1_{AC}$).
Any such observed value of the local EP can be plugged into \cref{eq:third_IFT}, to give a conditional IFT.
If now we allow the scientist to measure the work in the second solitary process, along
with the associated joint states, and to know the associated distributions $p^1$ and $p^2$,
then this theoretical prediction of  \cref{eq:third_IFT} will be confirmed.

Note that since both solitary processes are noisy, and $x_C$ is not reset to some
standard distribution between the running of the two solitary processes, the trajectories over
the two processes will not be independent. So there will be statistical coupling between $x_i(t)$   
and $x_j(t')$ for all times $t, t'$, for all nine choices of the associated pair of systems, $\{i, j\}$.

\section{Thermodynamic Uncertainty Relations for Bayes nets}
\label{app:TURs}

In this appendix I show how to combine Eq.\,(7) in the main text
with several TURs  to derive 
bounds on the precision of time-integrated currents in BNs. 

First, note that while the specification of the BN involves
discrete time, each solitary process transpires in continuous time.
(The BN only specifies the associated marginal state distributions at a sequence of discrete times
during the continuous-time process.) So
the relevant TURs are the ones for continuous time dynamics, \textit{not} the TURs for discrete time
dynamics~\cite{chiuchiu.pigolotti.discrete.time.TURs.2018,Proesmans.discrete.time.TUR_2017}.

It will be convenient to 
write the expected EP during some (implicit) time interval as $\sigma(\vP)$, with the density function $\vP$ explicit.
Recall that a real-valued function $J(\vx)$ is called a time-integrated \textbf{current} if it is
time-antisymmetric, i.e., if $J(\tilde{{\vx}}) = -J(\vx)$.
Since $\vx$ is a random variable, so is $J(\vx)$. The \textbf{precision} of $J$ is defined as 
$\langle J \rangle^2 \;/\; {\mbox{Var}} (J)$,
and measures the average size of the fluctuations in the value of $J(\vx)$. 

Note that there are an infinite
number of current functions $J(.)$. Nonetheless, recently there has been a flurry of
results in the literature that upper-bound the precision of \textit{every} current $J(\vx)$
with a ($J$-independent) function of $\langle \sigma \rangle$, the expected EP during the process that generates the trajectories
$\vx$~\cite{falasco2019unifying}.
These results  --- called ``thermodynamic uncertainty relations'' (TURs) --- mean that we
cannot increase the precision of any current beyond a certain point without paying for it by increasing EP.
Alternatively, they mean that if we can experimentally measure the precision of a current, then we can lower-bound the sum of all
contributions to $EP$ that are not directly experimentally measurable.

The TURs differ from one another in the restrictions they impose on $\vP$.
As an example, if the system is in a NESS when it runs, any associated
current $J(\vx)$ obeys the bound~\cite{gingrich_horowitz_finite_time_TUR_2017},
\eq{
\dfrac{2\langle J \rangle^2}{{\mbox{Var}} (J)} &\le \langle \sigma \rangle
\label{eq:51}
}
A weaker version of this bound applies when the distribution over states varies over time, so long as 
two conditions are still met. First, 
the starting and ending distributions must be identical. Second, the driving protocol
must be time-symmetric, i.e., both the trajectories of Hamiltonians and the trajectory
of rate matrices must be invariant if we replace all times $t$ with $t_f + t_0 - t$, where $t_0$ and $t_f$ are the
beginning and ending times of the process, respectively. Under such
circumstances,  the TUR \cref{eq:51} is replaced with
\eq{
\dfrac{2\langle J \rangle^2}{{\mbox{Var}} (J)}
			&\le e^{{\langle \sigma \rangle}} - 1
\label{eq:25}
}
which is known as the ``generalized thermodynamic uncertainty relation'' (GTUR~\cite{hasegawa2019generalized}).

More recently, a variant of the TURs was derived which upper-bounds the instantaneous
current at the end of the process rather than the integrated current across the entire process~\cite{liu2019thermodynamic}.
Suppose the process takes place during the time interval $[0, \tau]$. Write the ending instantaneous
current as $j_\tau(\vx) = \sum_{x\ne x'} W^{x'}_x(t) \vP(\vx'_t) d_{x',x}$, where $d_{x,x'}$ is any
antisymmetric matrix, $W(t)$ is the rate matrix of the underlying CTMC at time $t$, and $\vP(\vx'_t)$ is
the probability that the state at time $t$ is $\vx'_t$. 
Then this new TUR says that if the rate matrix and Hamiltonian are both constant
during the process (as in a NESS, but also more generally), then
\eq{
\dfrac{2 \tau \left\langle{j}_\tau \right\rangle^2}{{\mbox{Var}} (J)} &\le \langle \sigma \rangle
\label{eq:26aa}
}
Importantly, this bound holds regardless of the forms of the beginning and ending distributions.

In light of these results, suppose that
each separate solitary process in the BN is time-symmetric about the middle of the interval in
which that solitary process takes place. 
Assume as well that the beginning marginal distribution
of every solitary system $[v]$ when the solitary process associated with node $v$ begins to run is the same as the ending marginal distribution of that solitary
system after that solitary process finishes running. (Formally, this means that for all $v \in V$, $P(x^{v-1}_{[v]}) = P(x^v_{[v]})$.)
%

This second assumption means that the marginal distribution over the state
of any one subsystem $i$ in the BN, $P(x_i)$,
is the same at the beginning of the running of the entire BN as at the end of the running of the entire
BN. This is true even if that variable corresponds
to multiple nodes in the Bayes net, i.e., if $g^{-1}(i)$ contains more than one element of $V$, using
the terminology of \cref{app:BN_details}. However, for any node $v$ in the BN, in general 
the \emph{joint} distribution over the states of the subsystems $[v]$, $P(x_{g([v])})$, can differ arbitrarily between
the beginning and the end of the running of the entire BN. (That is because  each of the subsystems in $[v]$ can also change
state in the implementation of some other nodes of the Bayes net besides $v$.)

For this scenario, taking expectations of both sides of 
Eq.\,(7) in the main text
and then applying \cref{eq:25} gives
\eq{
\langle \sigma \rangle &=\sum_v \left( \langle{{\sigma}}_{[v]}\rangle) - \langle \Delta I^v\rangle_{\vP^v} \right) \nonumber \\
	&\ge \sum_v 	\left(  \ln \bigg[\dfrac{2 \langle J_{[v]} \rangle^2}{{\mbox{Var}}(J_{[v]})} +1\bigg] -  \langle \Delta I^v\rangle \right)  
\label{eq:26aaa}
}
where the random variable $J_{[v]}$ is any time-asymmetric function of the components
of the trajectory segment $\vx^v_{[v]}$. (Note that by definition of solitary process,
the only such function that can be nonzero must involve changes in the value of $x_v$.)
This establishes Eq.\,(12) 
in the main text.

In the special case that each subsystem $v$ is
actually in a NESS when it runs, we can apply \cref{eq:51} to establish the stronger bound,
\eq{
\langle \sigma \rangle 
	&\ge   \sum_v\left( \dfrac{2 \langle J_{[v]} \rangle^2}{{\mbox{Var}}(J_{[v]})} -  \langle \Delta I^v \rangle \right)  
\label{eq:26bb}
}
where the subscripts ${\vP^v_{[v]}}$ in the expectations and the variance are implicit. This
establishes Eq.\,(13)
in the main text.

Finally, if in fact the process is time-homogeneous, then no matter what the
beginning and ending distributions are, we can use \cref{eq:26aa} rather than \cref{eq:51}, to establish
\eq{
\langle \sigma \rangle 
	&\ge   \sum_v\left( \dfrac{2 \langle \tau_v j_{[v]}^{t_v}\rangle^2}{{\mbox{Var}}(J_{[v]})} -   \langle \Delta I^v \rangle \right)
\label{eq:26cc}
}
where the duration of the process updating node $v$ is $\tau_v$, $t_v$ is the time that that
process ends, $j_{[v]}^{t_v}$ is the instantaneous current over $X_{[v]}$ evaluated at $t_v$, and
$J_{[v]} = \int_{t_v - \tau_v}^{t_v} dt \, j_{[v]}^{t}$. This establishes Eq.\,(11) 
in the main text.

\cref{eq:26aaa,eq:26bb,eq:26cc} illustrate a trade-off among the precisions of (instantaneous) currents of the various 
subsystems, the sum of the drops
in mutual information, and the total dissipated work of the joint system. 

\begin{example}
Return to the NESS scenario considered at the end of 
Sec. IV in the main text.
Consider implementing that same BN with a different physical process. 
Just like the process described there, this alternative process would
first update the state of $X_A$, and then when that was done it would update the state of $X_C$. However, those two updates
would not be done with solitary processes. Instead, the state of $X_A$ would be updated with a CTMC whose rate matrix evolves $x_A$ based
on the current state of all three variables, $x_A, x_B$, and $x_C$. (In contrast, as discussed in \cref{app:path_wise_sub}),
if $x_{AB}$ runs with a solitary process, then the associated rate matrix can only involve $x_A$ and $x_B$.) 

This would allow the CTMC to exploit the initial coupling of
$x_A$ and $x_C$ in order to reduce the total EP that is generated by updating $x_A$. 
Similarly, the state of $X_C$ would be updated with a CTMC whose rate matrix evolves $x_C$ based
on the then-current state of all three variables, $x_A, x_B$, and $x_C$, and thereby reduce the total EP generated by updating $x_C$.
The end result is that Eq.\,(15) in the main text
would still hold, only with the $\Delta I^v$ terms removed. 
Since those terms are both negative, this would (in theory) allow the process to generate the same global EP as the original process but with greater precisions of both of the currents.
\end{example}

\end{document}